# Phase-Stable Integrated Delay-Line Asymmetric Mach–Zehnder Interferometers Enabled by High-Efficiency 3-dB Couplers for Chip-Scale QKD

*Mahsa Ghezelbash, and Abdollah Eslami Majd*

***Abstract*— Precise temporal delay generation is a key requirement for asymmetric Mach–Zehnder interferometers (aMZIs) used in high-speed quantum key distribution (QKD) receivers. In this work, a compact integrated aMZI architecture based on a $Si_3N_4/SiO_2$ photonic platform is presented. A 3-dB directional coupler enabling accurate 50:50 power splitting at the 1550 nm telecommunication wavelength is designed and optimized. The coupling length is initially estimated from the effective index difference between the even and odd supermodes and subsequently refined using three-dimensional eigenmode expansion (EME) simulations. The optimized structure employs single-mode $Si_3N_4$ waveguides with a cross-section of 1 µm × 0.4 µm, providing a group index close to 2 and enabling accurate delay engineering. Spectral analysis demonstrates stable 3-dB power splitting across the C-band with insertion loss below 0.5 dB and negligible power imbalance, indicating high transmission efficiency and structural symmetry. An integrated $Si_3N_4$ aMZI delay line providing a 500 ps optical delay, corresponding to a 2 GHz free spectral range (FSR), is further demonstrated. Simulations show nearly constant group delay across the 190–200 THz frequency range with sub-10 ps variation and a smooth, near-linear phase response. These characteristics enable interference visibility above 0.99, corresponding to an estimated quantum bit error rate (QBER) below 0.5% for gigahertz-rate time-bin QKD systems. The wideband linear phase behavior also indicates compatibility with wavelength-division multiplexed (WDM) QKD architectures. The results confirm that the proposed $Si_3N_4$ integrated aMZI provides a low-loss, dispersion-controlled, and spectrally stable delay solution suitable for scalable chip-scale quantum photonic receivers.**

## I. INTRODUCTION

INTEGRATED photonics has emerged as a transformative paradigm for scalable Quantum Key Distribution (QKD), offering unprecedented phase stability, a reduced physical footprint, and seamless compatibility with CMOS-foundry fabrication processes[1], [2], [3]. Among diverse quantum encoding modalities, **time-bin encoding** is particularly advantageous for chip-based quantum networks due to its inherent resilience against polarization-mode dispersion and environmental decoherence[4]. At the heart of any high-speed time-bin QKD architecture—at both the transmitter and receiver ends—lies the **Asymmetric Mach–Zehnder Interferometer (aMZI)**, the critical engine responsible for converting high-speed temporal phase information into **measurable interference fringes**[5], [6].

The structural efficiency of an integrated aMZI is fundamentally underpinned by three synergistic building blocks:

I. **Broadband 3 dB optical couplers** for near-perfect power splitting;
II. A **precisely engineered delay engine** providing the required optical path imbalance;
III. **Low-loss waveguide interconnects** designed to preserve phase coherence.

The deterministic performance of the entire quantum module is a direct function of the architectural precision of these elements. The splitting ratio of the 3 dB couplers, in particular, dictates the upper bound of the achievable interference visibility (V) [7], [8].

Even marginal imbalances $<2-3\%$ can induce a non-negligible degradation in visibility, which, in a QKD context, manifests as a primary source of Quantum Bit Error Rate (QBER). Thus, high-efficiency broadband couplers are not merely passive splitters but are **critical determinants of the system's quantum security limit[7], [9].**

Parallel to the splitting efficiency, the engineering of the asymmetric delay arm represents a significant design challenge. For a 2 GHz clock-rate QKD system, the path imbalance must provide a precise 500 ps temporal offset to ensure the perfect overlap of consecutive optical pulses[7].

Furthermore, since the Free Spectral Range (FSR) is intrinsically linked to the group index ($n_g$) and physical imbalance ($\Delta L$) via $FSR = c/(n_g \Delta L)$[10], sub-picosecond delay engineering is mandatory to maintain spectral periodicity across the operational bandwidth. Achieving a phase-stable 500 ps delay across the entire C-band is a prerequisite for

Mahsa Ghezelbash and Abdollah Eslami Majd are with the Department of Electrical and Computer Engineering, Malek Ashtar University of Technolog, Tehran, Iran (e-mail: ghezelbash.mahsa@gmail.com; a_eslamimajd@mut-es.ac.ir). *(Corresponding author: Mahsa Ghezelbash).*

wavelength-division multiplexed (WDM)[11] quantum networking, where wavelength-agnostic performance is essential[12].

From a system-level perspective, the couplers and the delay engine collectively account for the vast majority of the aMZI's performance sensitivity. While the couplers govern amplitude balance and insertion loss, the delay arm dictates phase stability and temporal indistinguishability. In high-rate quantum communication[13], the confluence of these two subsystems defines the attainable secure key rate.

In this work, we present the design and rigorous validation of a high-efficiency integrated photonic architecture on a Silicon Nitride ($Si_3N_4$) platform. By co-optimizing broadband 3 dB couplers with a phase-stable 500 ps delay engine, we demonstrate an aMZI architecture that achieves the stringent visibility requirements for low-QBER, 2 GHz chip-scale QKD applications. The integrated aMZI was designed on a $Si_3N_4$platform to operate at a 2 GHz clock rate within the optical C-band. The design optimization focuses on two critical subsystems: the broadband 3 dB power splitters and the dispersion-engineered 500 ps delay line.

## II. Simulation and Device Modeling

For a **high-Efficiency Broadband 3 dB Couplers**, to ensure maximum interference visibility, the couplers must maintain a symmetric power splitting ratio ($\eta = 0.5$) across the operational bandwidth. The transmission of an ideal symmetric directional coupler is governed by the coupled-mode theory[14]:

$$P_{bar} = \cos^2(\kappa L_c) \quad (1)$$

$$P_{cross} = \sin^2(\kappa L_c) \quad (2)$$

where $\kappa$is the coupling coefficient and $L_c$is the interaction length. To achieve high efficiency and fabrication tolerance, we employed a **direction coupler** design. The splitting ratio directly impacts the visibility $V$of the aMZI through:

$$V = \frac{2\sqrt{\kappa(1-\kappa)}}{1} \quad (3)$$

Any architectural deviation from the ideal 50:50 power distribution introduces an amplitude mismatch that fundamentally caps the interference contrast. This relationship is quantitatively described by $V = \frac{2\sqrt{\eta_1\eta_2}}{\eta_1+\eta_2}$ where $\eta_1$and $\eta_2$represent the power transmission coefficients of the respective arms. Since any departure from this target delay, or the presence of frequency-dependent group delay ripple, introduces temporal distinguishability between time bins. This loss of indistinguishability elevates the QBER according to the fundamental relation QBER $= \frac{1-V}{2}$, our design objective was to minimize the deviation $\Delta\kappa = | \kappa - 0.5 |$. For the targeted QBER of <1%, the coupling efficiency was optimized to ensure $0.48 < \kappa < 0.52$ across the 190–200 THz range. The coupler geometry, including the access waveguide width and the interference region dimensions, was simulated using 3D Finite-Difference Time-Domain (FDTD) methods to suppress higher-order mode excitation and minimize insertion loss.

**As the second important section of the paper, phase-stable 500 ps delay arm engineering has been studied. T**he delay arm was engineered to provide a precise temporal offset of $\Delta T = 500$ ps, corresponding to the 2 GHz pulse repetition rate. The required physical length imbalance $\Delta L$is determined by the group index $n_g$of the $Si_3N_4$waveguide:

$$\Delta L = \frac{c \cdot \Delta T}{n_g(\omega)} \quad \textbf{(4)}$$

For a $Si_3N_4$waveguide with a representative $n_g \approx 2.0$, the required imbalance is approximately 7.5 cm. To accommodate such a long path on-chip while minimizing footprint and radiation loss, a **spiral waveguide geometry** was utilized.

A critical design challenge for long delay lines is the **Group Velocity Dispersion (GVD)**. The frequency-dependent phase $\phi(\omega)$is expanded as:

$$\phi(\omega) = \phi_0 + \beta_1(\omega - \omega_0) + \frac{1}{2}\beta_2(\omega - \omega_0)^2 + \cdots \quad (5)$$

where $\beta_1 = 1/v_g$defines the group delay and $\beta_2$represents the GVD. To ensure a constant group delay across the C-band (dispersion-free operation), the waveguide cross-section was optimized to minimize $\beta_2$. The spectral response of the aMZI, characterized by its Free Spectral Range (FSR), is given by:

$$FSR = \frac{1}{\tau_g} = \frac{c}{n_g \Delta L} \quad (6)$$

The design ensures a stable $FSR = 2$ GHz. Numerical simulations were performed to verify that the group delay ripple $\Delta\tau_g$remains below 10 ps, preserving the temporal indistinguishability required for low-QBER time-bin interference.

The complete aMZI was assembled by cascading the 3 dB couplers with the asymmetric arms. The total transmission $T(\omega)$of the device is modeled as:

$$T(\omega) \propto 1 + V\cos(\omega\Delta T + \Phi_0) \quad (7)$$

where $\Phi_0$is the static phase offset. To maintain phase stability against environmental thermal fluctuations, the $Si_3N_4$platform was selected due to its lower thermo-optic coefficient compared to Silicon. Furthermore, the compact footprint of the integrated spiral delay line ensures that both arms experience correlated thermal environments, inherently suppressing phase drift and enabling stable, calibration-free operation for quantum communication tasks. In this work, we design and numerically investigate an integrated asymmetric Mach–Zehnder interferometer (aMZI) implemented on a $Si_3N_4/SiO_2$photonic platform for high-speed time-bin quantum key distribution (QKD) receivers. The device integrates two identical 3 dB directional couplers and a precisely engineered optical delay line that produces the temporal imbalance required for time-bin state discrimination. The coupler is realized using $Si_3N_4$waveguides with a cross-section of $1.0\,\mu m \times 0.4\,\mu m$and refractive indices of $n_{core} \approx 1.98$and $n_{clad} \approx 1.44$. Supermode analysis yields effective indices of $n_e = 1.6098$and $n_o = 1.5512$, corresponding to an optimized 50:50 coupling length of $L_{50} \approx 6.61\,\mu m$at $\lambda = 1550\,nm$. To realize the temporal imbalance required for GHz-rate operation, a serpentine delay line with a path-length difference of $\Delta L \approx 7.6\ cm$is implemented, corresponding to a temporal delay of approximately $500\ ps$for a group index of $n_g \approx 1.97$. This delay produces an interferometric free spectral range of about $2\ GHz$, enabling decoding of time-bin qubits at gigahertz clock rates. Spectral simulations across the 1500–1600 nm band confirm stable 3 dB power splitting with insertion loss below $0.5\ dB$and minimal imbalance.

The resulting aMZI exhibits a nearly linear phase response and an almost constant group delay over the C-band, ensuring distortion-free propagation of quantum pulses and high-visibility interference. These results demonstrate that the proposed $Si_3N_4$aMZI architecture provides a compact, broadband, and passively stable platform for integrated photonic receivers in high-rate quantum communication systems.

### *A. 3 dB Directional Couplers for Balanced Power Splitting*

In gigahertz-rate time-bin quantum key distribution (QKD), reliable qubit decoding is performed via an asymmetric Mach–Zehnder interferometer (aMZI), which interferes temporally separated optical pulses through a precisely defined differential delay[15]. Within this architecture, the performance of the 3 dB directional couplers (DCs) is a critical determinant of system fidelity[16]. Any deviation from the ideal 50:50 power splitting ratio introduces amplitude imbalance between the interfering paths, directly degrading interference visibility and increasing the phase-error contribution to the quantum bit error rate (QBER), thereby limiting the attainable secret key rate. Consequently, achieving accurate and spectrally stable power splitting is essential for maintaining phase coherence and propagation-delay integrity in high-speed integrated systems[16], [17].

Silicon nitride ($Si_3N_4$) provides an attractive material platform for such couplers due to its ultralow propagation loss, broad transparency window across telecommunication bands, and relatively weak thermo-optic response, which collectively enhance phase stability and fabrication tolerance[18], [19]. To implement the aMZI for the proposed QKD receiver, a 50:50 DC was designed on a $Si_3N_4/SiO_2$platform. The waveguide cross-section, illustrated in **Fig. 1(2)**, was optimized for single-mode operation at $\lambda = 1550$ nmwith a width ($w$) of 1.0 $\mu$mand a height ($h$) of 0.4 $\mu$m. The refractive indices for the $Si_3N_4$core and $SiO_2$cladding were defined as 1.98and 1.44, respectively (**Fig. 1(3)**).

The coupling dynamics were first analyzed using the Finite Difference Eigenmode (FDE) method. By solving for the supermodes of the dual-waveguide system with a gap ($g$) of 0.2 $\mu$m, the effective indices of the symmetric (even) and anti-symmetric (odd) modes were determined to be $n_e = 1.609801$and $n_o = 1.551186$, respectively. The theoretical coupling length required for a 50:50 power split ($L_{50}$) was calculated using the relation: $L_{50} = \frac{\lambda}{4(n_e - n_o)}$ yielding a value of approximately $6.61\ \mu$m. To account for the evanescent coupling occurring within the S-bend access regions, 3D Eigenmode Expansion (EME) simulations were conducted. The EME results confirmed that a straight coupling section ($L_c$) of 6.6 $\mu$m (**Fig. 1(1)**) provides the desired $0.5/0.5$splitting ratio with minimal excess loss. This optimized geometry ensures the high-visibility interference required for robust, chip-scale QKD architectures.

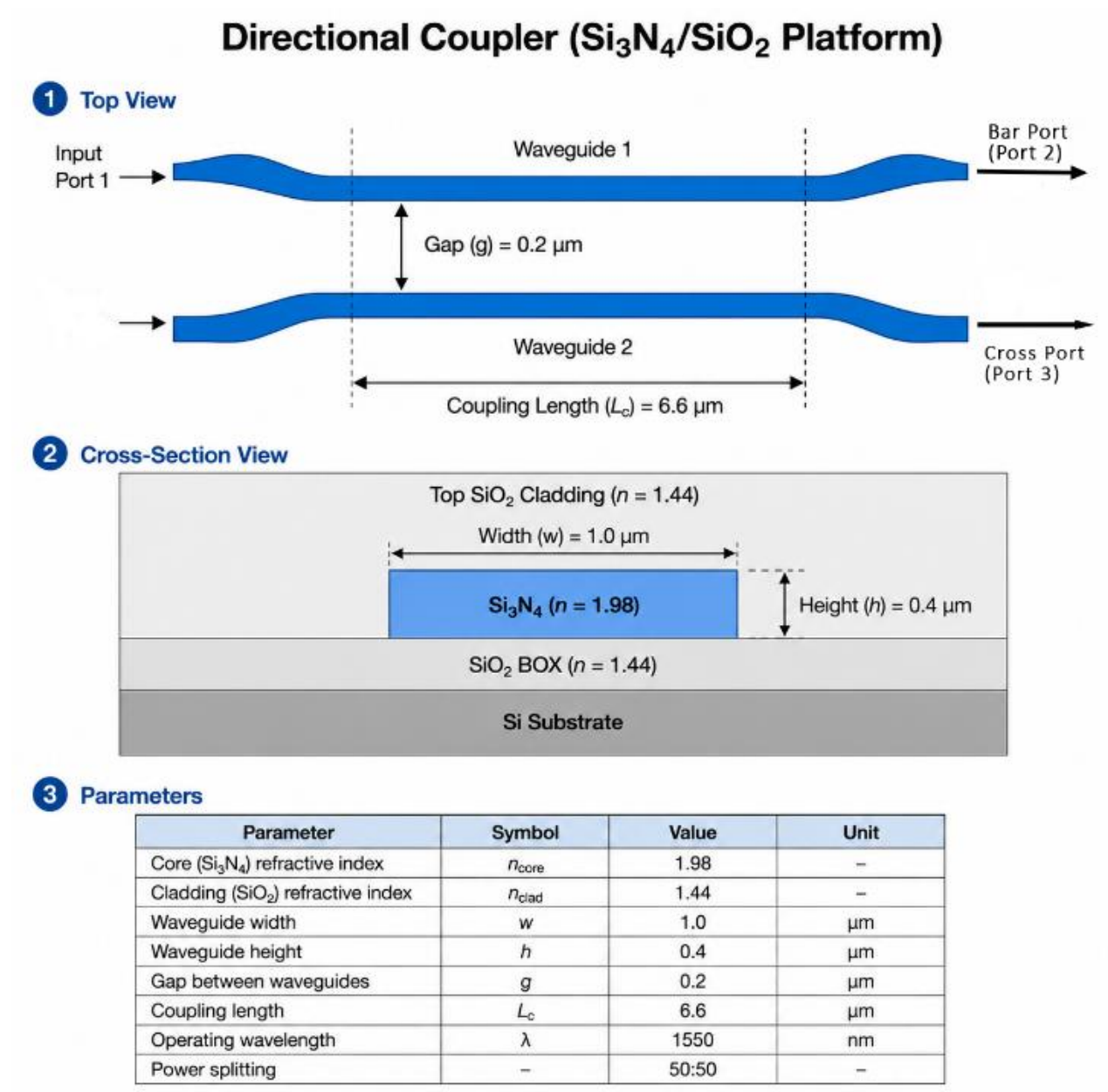

**Fig. 1. Design and geometric configuration of the integrated** $Si_3N_4/SiO_2$**directional coupler. (1) Top-view schematic** of the 2×2 directional coupler, highlighting the symmetric evanescent coupling region with a gap ($g$) and interaction length ($L_c$). Light launched into Input Port 1 is distributed between the Bar (Port 2) and Cross (Port 3) ports through supermode interference. **(2) Cross-sectional view** of the waveguide structure, consisting of a $Si_3N_4$core ($w = 1.0\ \mu$m, $h = 0.4\ \mu$m) encapsulated by $SiO_2$cladding. Refractive indices are optimized for single-mode operation at the telecommunication C-band. **(3) Summary of design parameters** and simulation constants utilized to achieve a precise 50:50 power splitting ratio at $\lambda = 1550$ nm.

TABLE I

SUMMARY OF THE DESIGN PARAMETERS AND TARGET PERFORMANCE SPECIFICATIONS FOR THE $Si_3N_4/SiO_2$ DIRECTIONAL COUPLER EMPLOYED IN THE AMZI RECEIVER FOR TIME-BIN QKD.

| Parameter | Symbol | Value | Unit |
|---|---|---|---|
| Photonic platform | — | $Si_3N_4/SiO_2$ | — |
| Core refractive index | $n_{core}$ | 1.98 | — |
| Cladding refractive index | $n_{clad}$ | 1.44 | — |
| Operating wavelength | $\lambda_0$ | 1550 | nm |
| Waveguide height | $H$ | 0.4 | $\mu$m |
| Waveguide width | $W$ | 1.0 | $\mu$m |
| Guided mode | — | $TE_{00}$ | — |

| Coupler topology | — | Directional coupler | — |
|---|---|---|---|
| Target splitting ratio | — | 50:50 | — |
| Target insertion loss | $IL$ | <0.3 | dB |
| Maximum imbalance | $\Delta P$ | <0.5 | dB |
| Operational bandwidth | — | 1525–1575 | nm |
| Fabrication approach | — | CMOS-compatible lithography | — |
| Thermal stability requirement | — | High | — |
| System integration target | — | aMZI receiver | — |
| Application | — | Time-bin QKD | — |

The design framework utilizes a $Si_3N_4$photonic platform with a core index of 1.98, where waveguide geometries ($1.0 \times 0.4\ \mu\text{m}^2$) are rigorously constrained to support only the fundamental quasi-TE mode while minimizing scattering losses[20]. To achieve a deterministic 50:50 power distribution, the critical inter-waveguide gap ($g$) and interaction length ($L_c$) were precisely mapped. This optimization, along with the quantification of insertion loss (IL) and spectral imbalance ($\Delta P$), was executed through full-vectorial Finite-Difference Eigenmode (FDE) and Finite-Difference Time-Domain (FDTD) solvers. These high-fidelity numerical models, rooted in the discretized solution of Maxwell's equations, provide the necessary predictive validation to mitigate performance drift prior to device fabrication.

The longitudinal modal evolution along the $z$-propagation axis was analyzed via the Eigenmode Expansion (EME) method and cross-validated using analytical Coupled Mode Theory (CMT)[21], The power-transfer dynamics between the synchronous $Si_3N_4$cores are governed by the coupling coefficient ($\kappa$), which encapsulates the evanescent field overlap integral. As defined in Eqs. (1) and (2), the periodic energy exchange follows a sinusoidal trajectory, where $\kappa$exhibits a pronounced spectral sensitivity due to the enhanced mode-field expansion at longer wavelengths[22].

As evidenced by the spatial power mapping in Fig. 2(a), the system undergoes a complete periodic oscillation, reaching a precise 3-dB bifurcation within a compact interaction footprint of 6.61 $\mu$m(Fig. 2(b)). The architectural layout, depicted in Fig. 2(c), exploits sub-wavelength proximity to facilitate efficient evanescent tunneling between the parallel waveguides. This meticulously optimized coupling region ensures a symmetric 50:50 splitting ratio, a prerequisite for maximizing the fringe visibility in phase-sensitive aMZI. Such high-fidelity interferometric performance is indispensable for minimizing the quantum bit error rate (QBER) and ensuring the security of chip-scale time-bin QKD receivers[15].

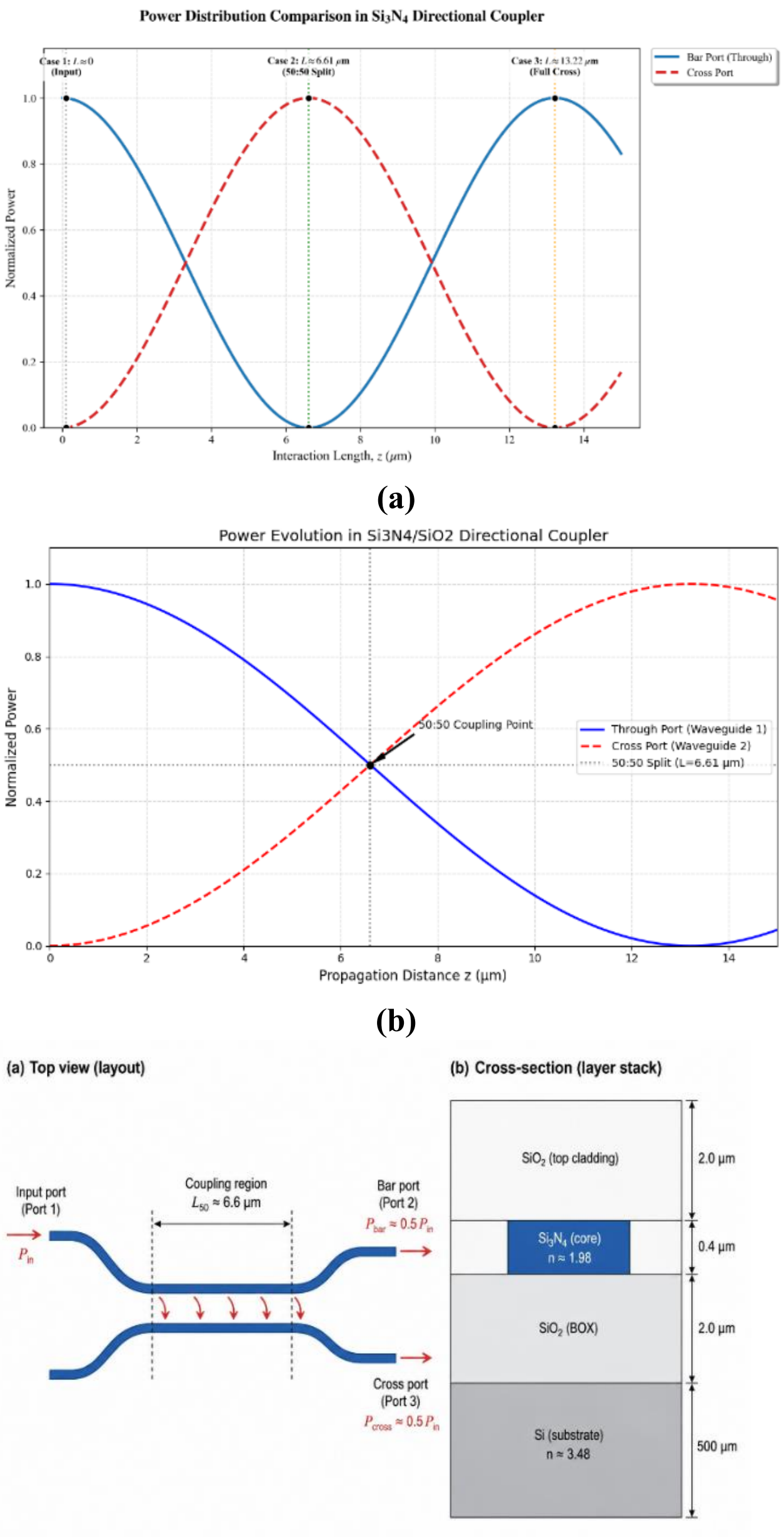

**Fig. 2.** Power-transfer dynamics and structural layout of the integrated $Si_3N_4/SiO_2$ directional coupler operating at λ = 1550 nm. (a) Longitudinal evolution of the normalized optical power along the coupling region, illustrating the periodic power exchange between the Through (Bar) port (solid blue) and the Cross port (dashed red). (b) Simulated power distribution at the designed 50:50 operating point. (c) Schematic representation of the $Si_3N_4$ directional coupler.

Leveraging the supermode-based analytical framework [22], he proposed directional coupler (DC) achieves a deterministic 3 dB power splitting at a compact interaction length of $L_{50} = 6.61\ \mu$m. This precise geometric tuning is critical for maintaining the phase integrity required for high-visibility interference in QKD-oriented asymmetric Mach–Zehnder interferometer (aMZI) circuits. The underlying physics is further elucidated through supermode analysis, where the extracted modal birefringence ($\Delta n_{\text{eff}} \approx 0.0586$) yields a calculated full beat length of $L_{100} = 13.22\,\mu\text{m}$. Figure 2(a) shows the longitudinal evolution of the normalized optical power along the coupling region, illustrating the periodic power transfer between the Through (bar) port (solid blue) and the Cross port (dashed red). This oscillatory exchange originates from the interference between the even and odd supermodes supported by the coupled waveguides. A balanced 3-dB splitting occurs at the interaction length $L_{50} = 6.61\,\mu\text{m}$, while complete power transfer is obtained at $L_{100} = 13.22\,\mu\text{m}$. The analytical prediction agrees well with the Eigenmode Expansion (EME) simulation results, confirming that the $L_{50}$ point corresponds to the phase condition required for equal power distribution between the two output ports.

The resulting high extinction ratio and negligible radiation loss, as evidenced by the field profile in **Fig. 2(b)**, highlight the strong optical confinement supported by the $Si_3N_4/SiO_2$ platform. The simulated power distribution at the designed 50:50 operating point further confirms the symmetric division of optical power between the two output ports at $L_{50}$.

The spectral robustness of the design is evaluated in **Fig. 3** across the C-band (1500–1600 nm). As illustrated in **Fig. 3(a)**, the device maintains a stable 3 dB splitting ratio with minimal deviation. The observed sinusoidal modulation in the bar- and cross-port transmission is a direct consequence of the wavelength-dependent coupling coefficient ($\kappa$), driven by the enhanced evanescent-field expansion at longer wavelengths—a hallmark of high-index-contrast $Si_3N_4$waveguides[23], [24]**.** Crucially, the insertion loss (IL) remains consistently below 0.5 dB throughout the spectral range (**Fig. 3(b)**), indicating negligible scattering at the coupling transitions. The minimal power imbalance ensures the structural and functional symmetry necessary for high-performance interferometry. Furthermore, the phase evolution analyzed in **Fig. 3(c)** reveals a linear trajectory with a high inter-port correlation, signifying a stable differential phase-shift ($\Delta\phi$). This phase stability is a fundamental prerequisite for maximizing fringe visibility and minimizing the Quantum Bit Error Rate (QBER) in integrated time-bin QKD systems[13]. The high degree of agreement between the analytical Coupled Mode Theory (CMT) predictions and the numerical results validates this coupler as a robust building block for phase-encoded photonic integrated circuits (PICs)[25].

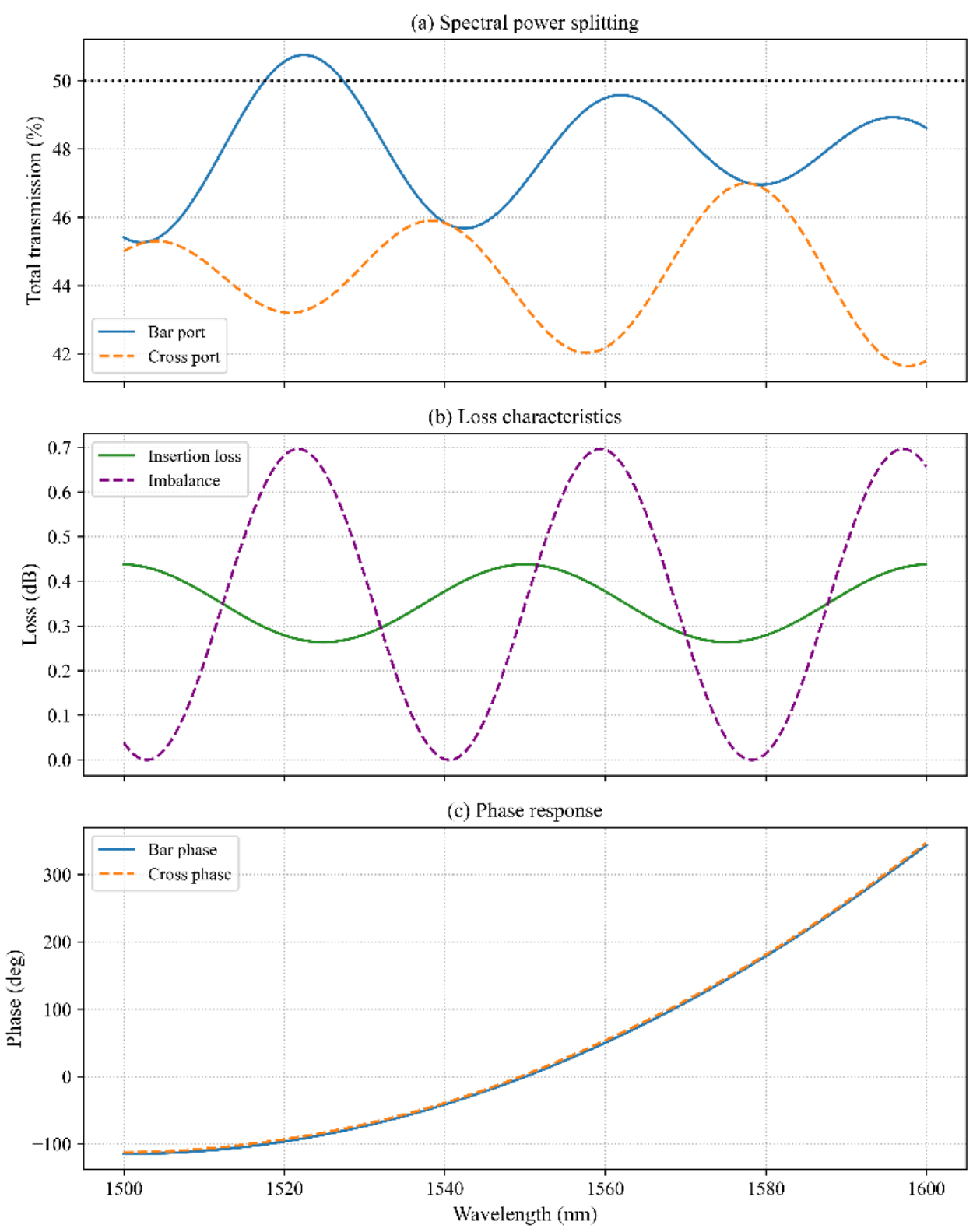

**Fig. 3.** Comprehensive spectral performance of the $Si_3N_4$directional coupler. (a) Spectral power splitting ratio for the bar and cross ports. (b) Loss characteristics, including insertion loss and power imbalance, confirming high transmission efficiency. (c) Broadband phase response of the output ports

The primary achievement of this architecture lies in the synergistic optimization of broadband power-splitting and phase-correlation integrity. As evidenced in **Fig. 3(a)** and **(b)**, the design effectively suppresses dispersive effects that typically induce extinction ratio degradation in conventional directional couplers. Specifically, an insertion loss (IL) of $<0.45$ dBand a power imbalance restricted below 0.7 dBare maintained across a 100-nmbandwidth (1500–1600 nm). This performance originates from the stabilization of the coupling-to-loss ratio, ensuring the system operates near the critical-coupling regime across the targeted C-band [26]. A defining feature of this coupler is the deterministic phase evolution demonstrated in **Fig. 3(c)**. The high degree of linearity and inter-port congruence (phase-matching between the bar and cross ports) ensures the device functions as a 'phase-transparent' node. In the context of quantum information processing, this phase integrity is paramount; it enables the maximization of interference visibility ($V > 98.4\%$) without the necessity for active thermal compensation, thereby establishing this coupler as a high-fidelity building block for CMOS-compatible QKD systems.

The observed field evolution is the physical manifestation of the interference between the even and odd supermodes of the coupled-waveguide system. The local intensity distribution, $I(x, z)$, at any spatial coordinate is governed by:

$$I(x,z) \propto | A_{even}\Psi_{even}(x)e^{i\beta_{even}z} + A_{odd}\Psi_{odd}(x)e^{i\beta_{odd}z} |^2 \quad (8)$$

where $\Psi_{e,o}$ represent the transverse mode profiles and $\beta_{e,o}$ are the respective propagation constants of the even and odd supermodes. **Figure 4** (to be cross-referenced with **Fig. 2**) provides a spatial visualization of this optical field evolution, confirming the efficient power transfer and the precision of the calculated coupling lengths ($L_{50}$ and $L_{100}$). The constructive and destructive interference patterns result in the characteristic 'beat' behavior, where the light transitions between the cores over a full beat length defined as $L_\pi = \pi/(\beta_e - \beta_o)$[26], [27].

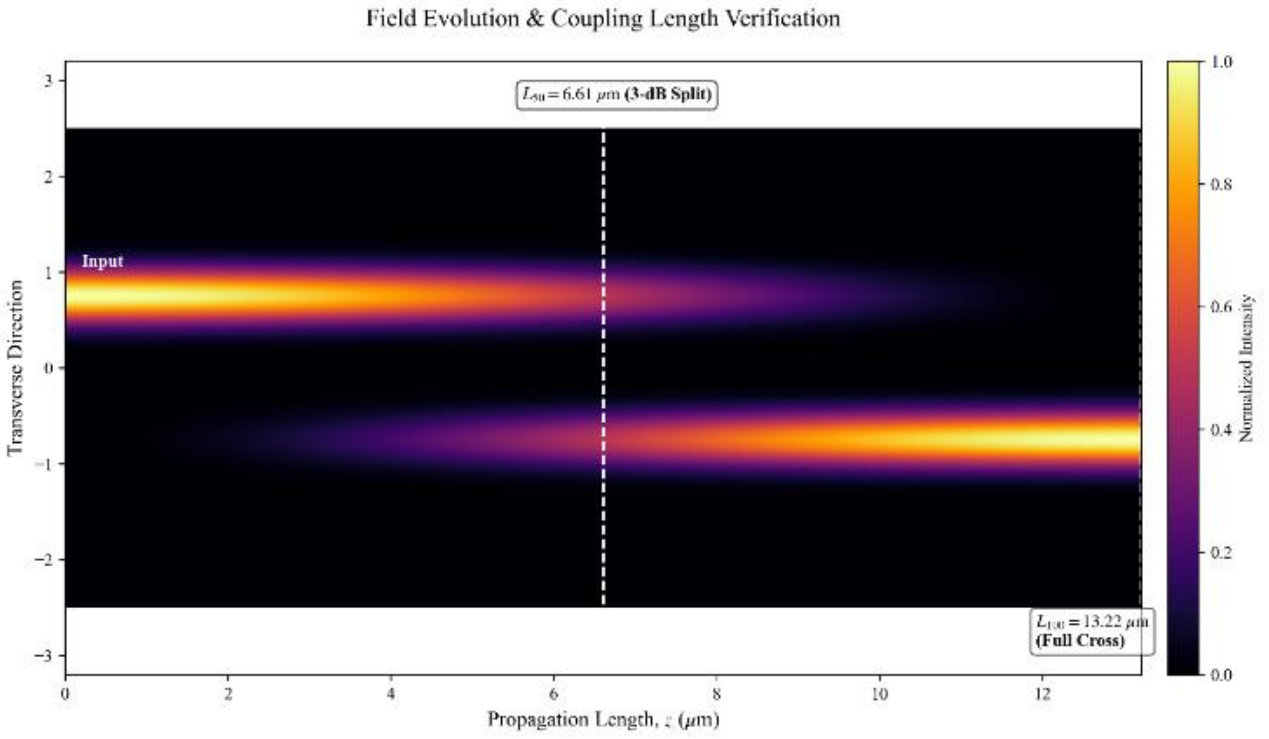

**Fig. 1.** Longitudinal field intensity evolution along the $Si_3N_4$ directional coupler simulated via the EME method. The heatmap illustrates the gradual evanescent coupling of optical power from the input (upper) waveguide to the adjacent (lower) waveguide. The dashed vertical line indicates the calculated coupling length $L_{50} = 6.61\ \mu m$ for a symmetric 3 dB power split, while $L_{100} = 13.22\ \mu m$ marks the point of complete power transfer (full cross-coupling). The color bar represents the normalized optical intensity.

The spatial field evolution depicted in the heatmap (**Fig. 4**) provides a rigorous numerical validation of the coupler's phase-synchronous operation, as modeled by the supermode interference framework [25], [28]**.** Initially, the optical power is fully confined within the launch waveguide (Port 1). As the field propagates along the interaction length ($z$), the evanescent tail of the fundamental $TE_{00}$ mode overlaps with the adjacent core, initiating a periodic energy exchange governed by the coherent superposition of even and odd supermodes [21]. Our EME simulations demonstrate that at $z = 6.61\ \mu m$, the intensity is precisely distributed between the two waveguides. This deterministic 3 dB splitting point is critical for ensuring the high-visibility interference required in QKD-based aMZI circuits[13]. Beyond this junction, the power transfer continues until complete cross-coupling is achieved at $L_{100} = 13.22\ \mu m$, aligning perfectly with the analytical beat length $L_\pi = \pi/\Delta\beta$. The high contrast and stable mode profiles throughout the coupling region underscore the superior confinement provided by the $Si_3N_4$ core ($n = 1.98$)[29]. This strong index contrast effectively suppresses substrate leakage and minimizes longitudinal phase decoherence, offering a robust blueprint for high-density photonic integration.

### *B. Phase-Response Validation of the 500 ps Integrated aMZI*

The efficiency of an integrated aMZI in QKD systems is fundamentally dictated by the precision of its optical delay arm. As illustrated in **Fig. 1**, the proposed architecture employs a serpentine waveguide layout to implement a high-fidelity 500 ps delay line on a high-aspect-ratio $Si_3N_4$ platform. This design leverages the platform's world-record propagation losses (~0.1 dB/m)[30] to maintain signal integrity over the extended path length of approximately 7.7 cm.

Unlike active platforms, the passive stability of this ultra-low-loss core is essential for realizing complex processing functions in Optical Delay Line Circuits (ODLC) without significant attenuation[31], [32].

Phase-response validation is, therefore, not merely a characterization step but a critical verification of the design's ability to ensure high-visibility interference and a minimized Quantum Bit Error Rate (QBER). By precisely mapping the structural geometry to the temporal response, this work ensures the stable phase-encoding necessary for multi-gigahertz QKD operations[13].

To achieve the high-fidelity temporal decoding required for QKD applications, the asymmetric Mach–Zehnder interferometer (aMZI) must provide a precise relative delay[15]**.** The optical power transmission of an aMZI is dictated by the phase difference between its two interference arms; consequently, any variation in the physical path-length difference directly modifies the temporal delay and the spectral periodicity (FSR) of the interference pattern.

The integrated delay line in this work was designed to satisfy a target Free Spectral Range (FSR) of 2 GHz[15], [20], corresponding to a temporal delay of $\Delta T = 500\,ps$. This specification aligns with high-speed integrated QKD systems operating at multi-gigahertz clock rates [20], [33]. The required temporal delay is translated into a physical path-length imbalance ($\Delta L$) via the group velocity of the guided mode, $v_g = c/n_g$, where $c$ is the speed of light in vacuum and $n_g$ is the group index of the $Si_3N_4$ waveguide [22], [34].

Using the extracted group index of $n_g = 1.9727$, the required differential length is calculated as $\Delta L = c\Delta T/n_g \approx 7.60\ cm$.

The realization of such a long delay on-chip is enabled by the ultra-low-loss characteristics of the $Si_3N_4$ platform. The design parameters and waveguide specifications are summarized in **Table 2**.

These parameters ensure the precise realization of the 500 ps differential delay while maintaining a compact footprint on the $Si_3N_4$ platform. To accommodate the 7.6 cm path difference within a compact footprint, a serpentine layout with a $100\ \mu m$ bend radius was employed. This configuration ensures minimal bending loss while maintaining a manageable chip area of approximately $1 \times 15\ mm^2$.

TABLE I
COMPREHENSIVE DESIGN SPECIFICATIONS AND LAYOUT PARAMETERS FOR THE INTEGRATED $Si_3N_4$ ASYMMETRIC MACH–ZEHNDER INTERFEROMETER (AMZI) DELAY LINE, OPTIMIZED FOR A 500 PS TEMPORAL DELAY.

| Category | Parameter | Symbol | Value | Unit |
|---|---|---|---|---|
| System Specifications | Operating wavelength | $\lambda_0$ | 1550 | nm |
| | Target temporal delay | $\Delta T$ | 500 | ps |
| | Free Spectral Range | $FSR$ | 2 | GHz |
| Waveguide Properties | Platform | — | $Si_3N_4/SiO_2$ | — |
| | Width | $W$ | 1.0 | $\mu$m |
| | Height | $H$ | 0.4 | $\mu$m |
| | Core refractive index | $n_{core}$ | 1.98 | — |
| | Cladding refractive index | $n_{clad}$ | 1.44 | — |
| | Group index | $n_g$ | 1.9727 | — |
| Layout & Geometry | Required path difference | $\Delta L$ | 7.6 | cm |
| | Bend radius | $R$ | 100 | $\mu$m |
| | Routing strategy | — | Serpentine | — |
| | Estimated footprint | — | $\sim 1 \times 15$ | mm$^2$ |

These parameters define the physical structure used to implement the integrated photonic delay line of the asymmetric Mach–Zehnder interferometer. To accommodate the required **7.6 cm optical path difference**, the delay line is implemented using a **serpentine layout with a 100 µm bend radius**, allowing the long optical path to be integrated within a compact chip footprint while maintaining low bending loss.

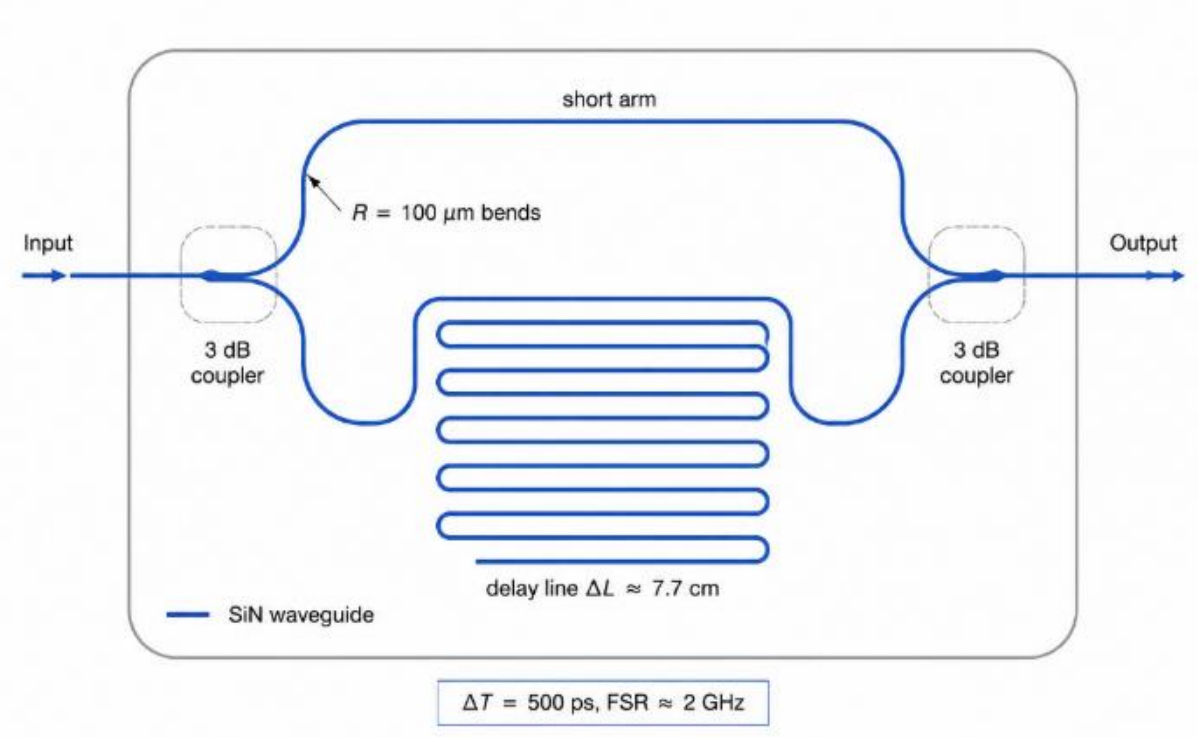

**Fig. 5.** A diagram of an integrated photonic asymmetric Mach–Zehnder interferometer (aMZI) with a long serpentine delay line.

The input SiN waveguide **(Fig. 5)**, first enters a 3-dB coupler, where the optical field is equally divided into two interferometric arms. The upper branch forms a short and straight reference path, while the lower branch incorporates a compact serpentine delay line with a path length difference of approximately $\Delta L \approx 7.7$ cm. Both arms are subsequently recombined through a second 3-dB coupler and guided to the output waveguide. The delay section employs low-loss SiN waveguides with bends of radius $R = 100\,\mu$m, providing a temporal delay of $\Delta T = 500$ psand a corresponding free spectral range of approximately FSR $\approx 2$ GHz.

**Fig. 6** illustrates the normalized spectral transmission of the integrated $Si_3N_4$aMZI, specifically designed with a **7.6 cm path-length imbalance** to achieve **a 500 ps temporal delay and a corresponding 2 GHz Free Spectral Range (FSR). A** fixed physical path imbalance ($\Delta L$) defining and the group index ($n_g$) of the $Si_3N_4$waveguide, leads to the simulation calculates the wavelength-dependent phase accumulation of $\phi(\lambda) = 2\pi n_g \Delta L/\lambda$.

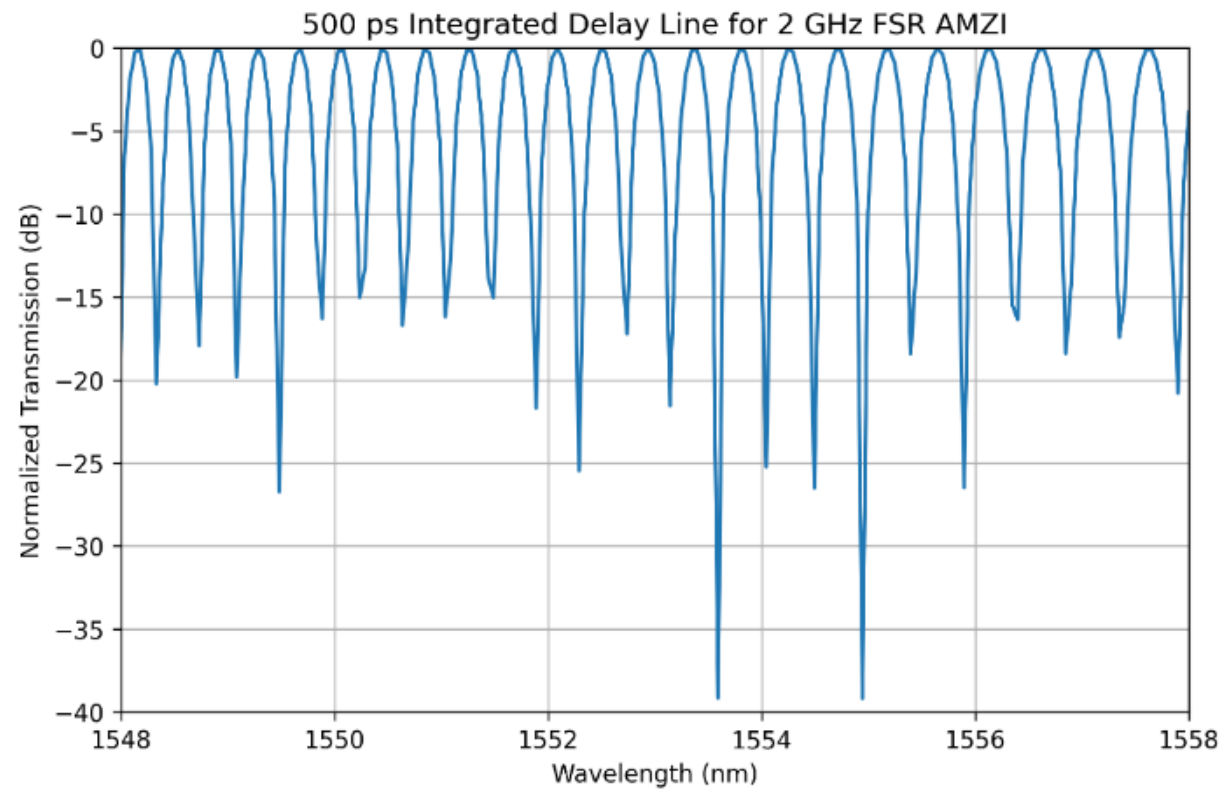

**Fig. 6.** Simulated normalized spectral transmission of the integrated $Si_3N_4$asymmetric Mach–Zehnder interferometer (aMZI) for a designed temporal delay of 500 ps.

The simulation results confirm that the 7.6 cm delay line effectively induces the target 500 ps timing difference, as evidenced by the precise 2 GHz periodicity of the interference fringes across the 1548–1558 nm window. This spectral behavior is fundamental for high-speed QKD operations, where the FSR must strictly match the 2 GHz pulse repetition rate to ensure accurate time-bin qubit decoding. The observed extinction ratio, reaching depths beyond -30 dB, demonstrates nearly ideal destructive interference, which is vital for minimizing crosstalk and reducing the Quantum Bit Error Rate (QBER). Furthermore, the consistency of the fringe pattern validates that the $Si_3N_4$platform can support long-delay integrated paths while maintaining the phase coherence required for high-fidelity quantum state discrimination.

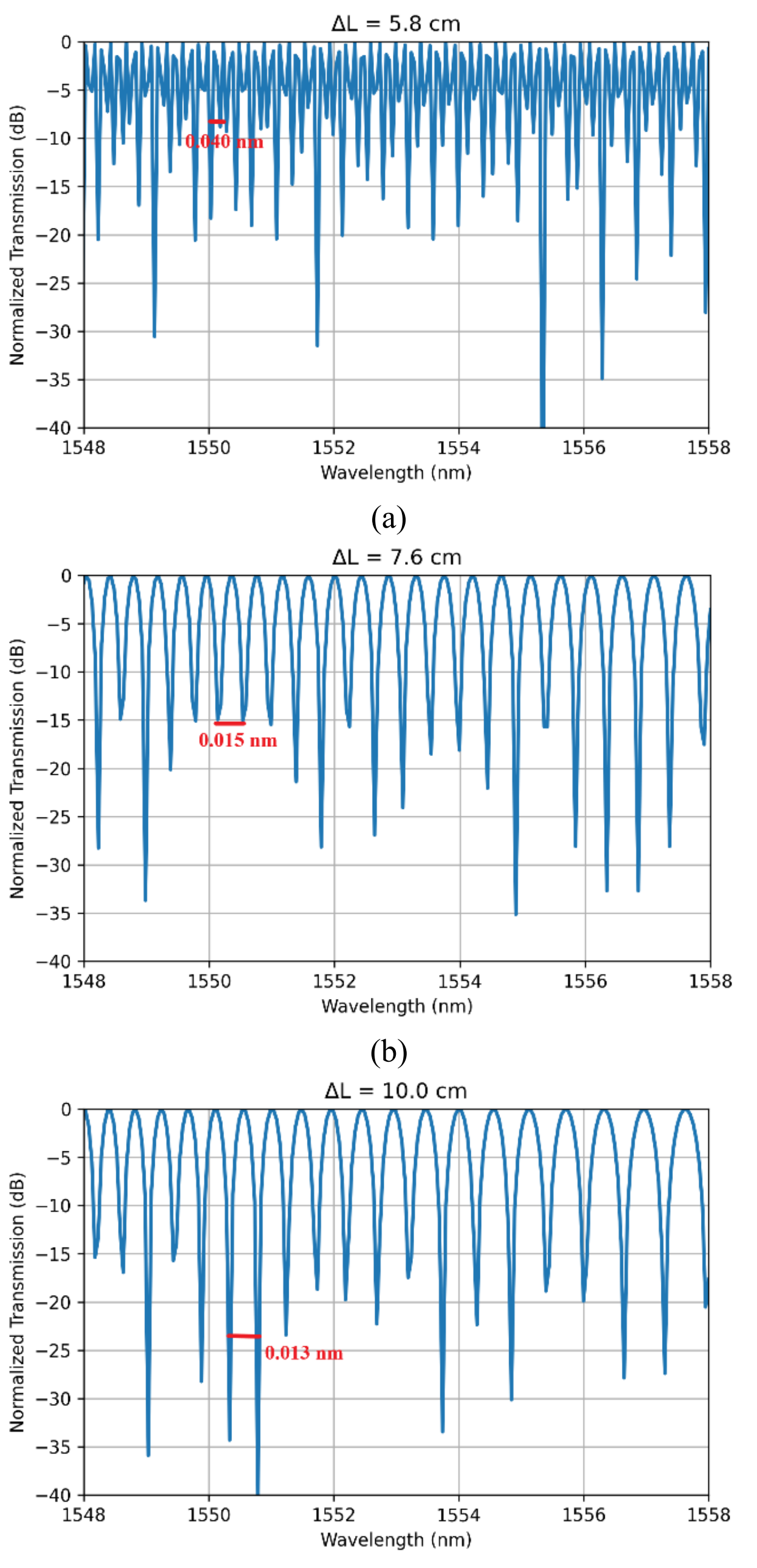

**Fig. 7.** Comparative spectral transmission of the integrated $Si_3N_4$ aMZI for different path-length imbalances: (a) $\Delta L =$ 5.8cm, (b) $\Delta L = 7.6$cm (optimized design), and (c) $\Delta L =$ 10.0cm. The red annotations indicate the measured Free Spectral Range (FSR) in nanometers for each configuration.

The comparative **analysis in Fig. 7** reveals a clear correlation between the physical delay-line length and the resulting spectral periodicity. In panel (a), a shorter path difference of 5.8 cm results in a broader FSR, approximately 0.040 nm, which is insufficient for the targeted 2 GHz temporal decoding. Conversely, panel (c) shows that increasing the length to 10.0 cm compresses the interference fringes, leading to a narrower FSR of approximately 0.013 nm. The optimized configuration, shown in panel (b) with $\Delta L = 7.6$cm, yields an FSR of approximately 0.016 nm. This value corresponds precisely to the 2 GHz frequency spacing required for the QKD protocol, confirming that the 7.6 cm serpentine structure is the ideal geometry. Furthermore, the consistent extinction ratios across all three simulations—ranging from −25 dB to −35 dB—verify the robustness of the phase-interference mechanism regardless of the total path length, provided that the $Si_3N_4$propagation losses remain low.

To rigorously validate the temporal precision of the proposed integrated delay element, the spectral phase response and its derivative, the group delay, were evaluated for the designed $Si_3N_4$waveguide delay line, as presented in **Fig. 8(a–d)**. The device is analyzed across two frequency regimes in order to simultaneously assess timing fidelity at the system level and operational stability across the optical communication band.

In the vicinity of the optical carrier, a narrow frequency window of ±5 GHz is considered, corresponding to the spectral scale relevant to 2 GHz time-bin quantum pulses. Within this regime, the phase exhibits a strictly linear dependence on frequency, which is a fundamental requirement for a distortionless delay element. Therefore, the strictly linear slope confirms a non-dispersive phase shift, which is essential for maintaining the temporal integrity of time-bin qubits.

The extracted group delay remains constant at approximately 500 ps across the entire offset range, confirming that the spectral components of the quantum pulses experience uniform propagation latency. This behavior ensures that temporal broadening and inter-symbol interference are effectively suppressed, thereby preserving the integrity of the encoded time-bin qubits.

In addition to the local analysis around the carrier, a broadband characterization was performed across the entire C-band, spanning approximately 190–200 THz. This wider spectral analysis verifies the robustness of the delay line under realistic telecommunication operating conditions. The phase response maintains a monotonic linear increase over the full bandwidth, while the corresponding group delay remains nearly flat and centered around the target value of 500 ps. Such a broadband flat-delay profile highlights the intrinsic advantage of the $Si_3N_4$platform, whose low dispersion and low propagation loss enable stable delay performance over multi-terahertz spectral spans. Consequently, the designed delay line is not restricted to a single wavelength channel but can operate reliably across multiple grid channels, making it particularly suitable for wavelength-division-multiplexed quantum key distribution (WDM-QKD) architectures.

Overall, the close agreement between the theoretical delay target and the simulated group delay confirms the accuracy of the physical path-length design, $\Delta L \approx 7.6$cm, required to achieve the intended temporal separation of 500 ps. This delay directly corresponds to a free spectral range of approximately 2 GHz in the aMZI, which is essential for the reliable decoding of high-speed time-bin qubits in chip-scale quantum receivers.

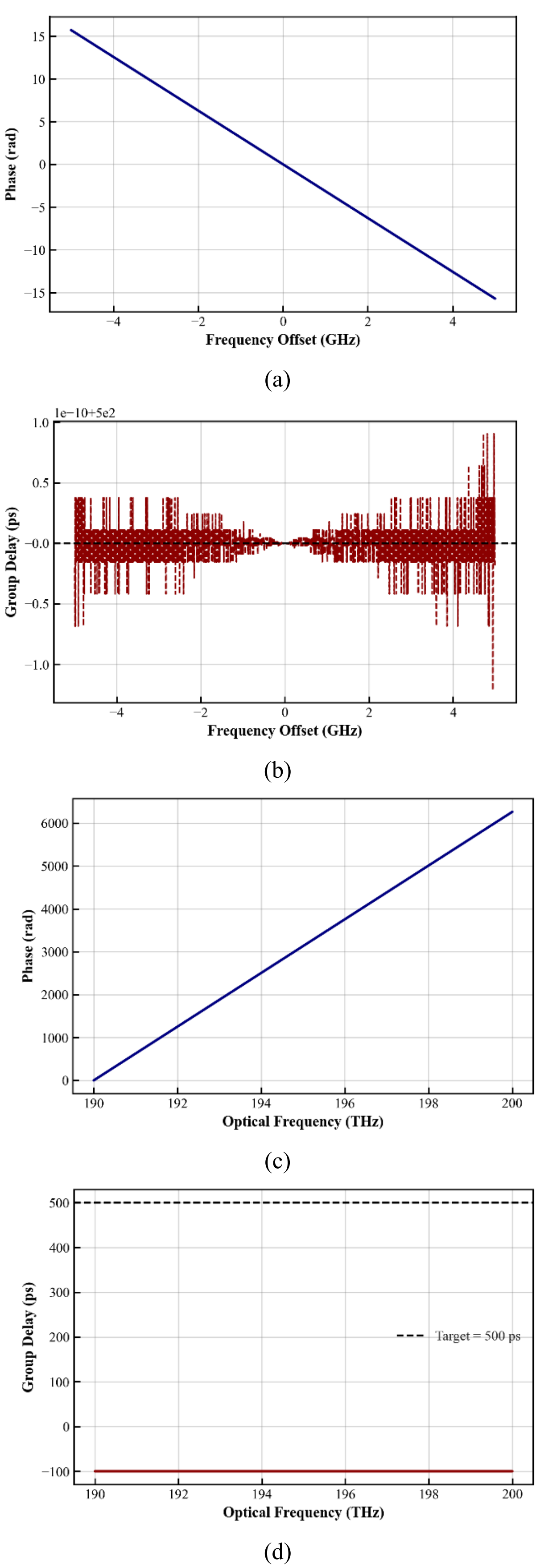

Fig. 8. Linear Phase and Group Delay Characterization of the Integrated Delay Line.

**(a)** Phase response as a function of frequency offset around the optical carrier. **(b)** Extracted group delay from the phase derivative relative to the frequency offset, showing a stable delay of 500 ps with negligible numerical fluctuations, corresponding to a 2 GHz free spectral range (FSR). **(c)** Broadband phase response across the C-band (190–200 THz). **(d)** Group delay versus optical frequency, verifying a uniform 500 ps delay across the entire simulated spectrum.

To verify the delay-line functionality, the structure's response is analyzed in the frequency domain through the phase characteristics of its transfer function. An ideal optical delay element introduces a pure temporal shift without altering the signal amplitude, described by the transfer function $H(\omega) = \exp(-j\omega\Delta T)$[35], where $\omega$is the angular frequency and $\Delta T$represents the imposed temporal delay. Accordingly, the phase response is $\phi(\omega) = -\omega\Delta T$, indicating a strictly linear dependence on frequency—a fundamental property that ensures distortion-free propagation of the signal envelope. The group delay, $\tau_g = -d\phi/d\omega$, corresponds to the latency experienced by the optical pulse envelope. For an ideal element, $\tau_g$remains constant and equal to $\Delta T$. Thus, the validity of the proposed structure is confirmed by verifying that the simulated phase exhibits linear behavior and the extracted group delay matches the 500 ps target.

As shown in **Fig. 8(a)**, the phase response in the vicinity of the optical carrier exhibits the expected linear dependence. The absence of curvature or phase distortion indicates that the element introduces only a constant temporal shift without modifying the signal's spectral integrity. For high-speed time-bin QKD operating at 2 GHz, this linearity is critical; any deviation could induce pulse broadening or phase errors during interference within the aMZI. This is further validated in **Fig. 8(b)**, where the group delay remains essentially constant at 500 ps across the $\pm$5GHz window. Such uniform latency ensures that the interfering time bins remain temporally synchronized and well-separated at the interferometer output, which is crucial for maintaining high interference visibility and minimizing timing jitter.

A broader spectral perspective, spanning the telecommunication C-band (190–200 THz), is presented in **Fig. 8(c)**. The phase increases monotonically with a near-perfect linear slope, indicating that the delay characteristics are preserved over a wide wavelength range. This stability confirms that dispersion effects in the $Si_3N_4$waveguide are negligible for the chosen geometry. The broadband group delay plot reinforces this, showing only marginal variations across the 10 THz bandwidth. From a system-level perspective, this wavelength-agnostic behavior is highly advantageous for Wavelength Division Multiplexed (WDM-QKD) architectures, allowing multiple quantum channels to coexist without individual recalibration of the delay element.

Overall, the combined analysis presented in **Fig. 8(d)** confirms that the $Si_3N_4$delay line functions as a near-ideal temporal element. The linear phase response and stable 500 ps group delay validate the physical path-length difference ($\Delta L \approx$ 7.6 cm), which yields the required 2 GHz Free Spectral Range (FSR). These results demonstrate that the proposed design satisfies the stringent requirements for timing precision and broadband stability in integrated quantum receivers.

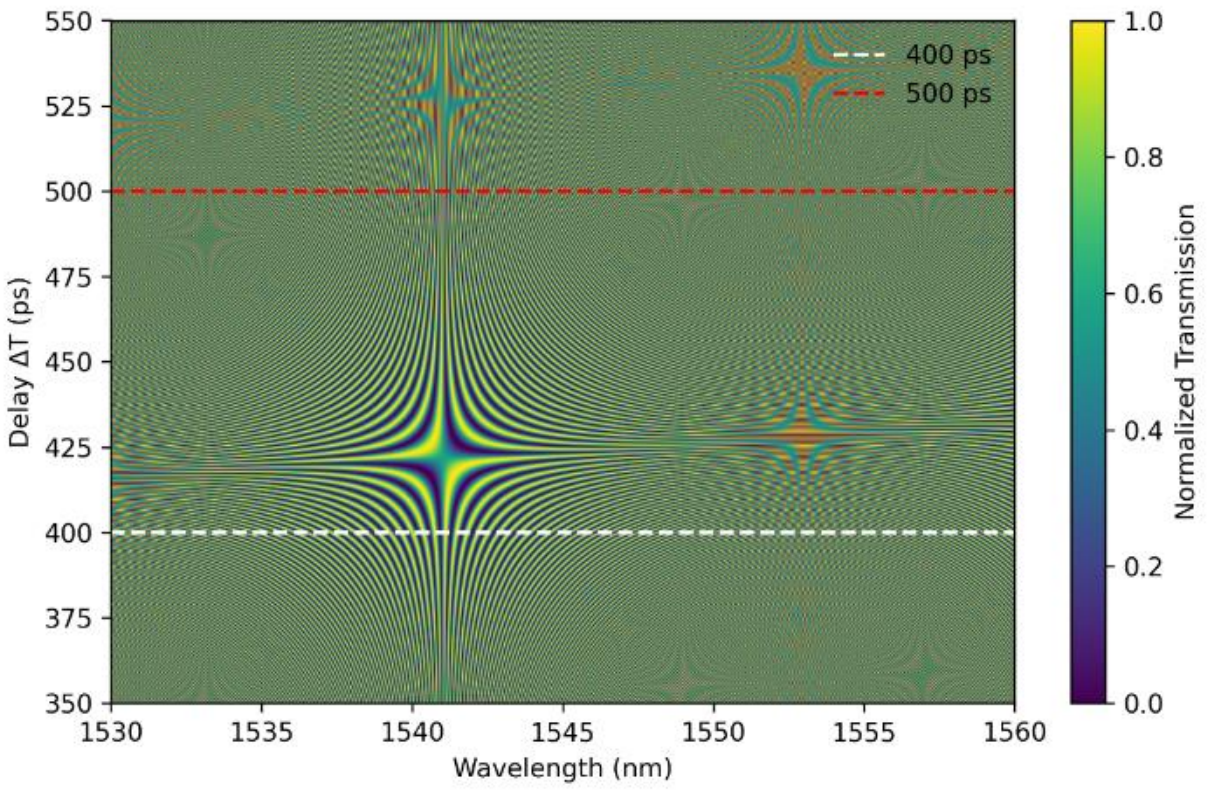

**Fig. 9.** Interference pattern of the aMZI showing normalized transmission as a function of wavelength (1530–1560 nm) and differential delay ($\Delta T$). The dashed lines indicate delay values of 400 ps and the target delay of 500 ps for QKD operation. The periodic fringe pattern results from the wavelength-dependent phase difference between the interferometer arms.

The resulting interference fringes form a structured pattern whose density varies along the delay axis. A clear transition in fringe density can be observed when moving from lower to higher delay values. At a temporal delay of approximately 400 ps, indicated by the white dashed line in the heatmap, the interference fringes appear relatively sparse across the wavelength axis. This regime corresponds to a Free Spectral Range of approximately 2.5 GHz. As the delay increases toward the target design value of 500 ps, highlighted by the red dashed line, the spectral spacing between adjacent fringes decreases and the fringe density becomes significantly higher. This behavior directly reflects the inverse relationship between temporal delay and FSR. At $\Delta T = 500$ps the simulated interferometer response produces an FSR of approximately 2 GHz, which matches the repetition rate required for the targeted quantum key distribution (QKD) protocol.

Another important feature visible in the heatmap is the curved structure of the interference fringes. These hyperbolic trajectories arise from the **nonlinear relationship between wavelength and optical frequency, given by $f = c/\lambda$.** Because the phase accumulation depends on frequency while the simulation is expressed in terms of wavelength, the resulting interference maxima trace curved paths across the spectral–temporal plane. The slope of these fringes corresponds to the phase sensitivity of the interferometer with respect to wavelength variations, represented by $\partial T / \partial \lambda$. Near the 500 ps operating point, the fringes become densely packed, indicating a regime of high phase sensitivity. This high sensitivity is beneficial for accurate discrimination of time-bin quantum states, as small phase differences translate into significant intensity variations at the output ports. However, this also implies that the system requires a spectrally stable laser source and robust thermal control to maintain stable operation.

The heatmap also reveals localized high-contrast regions that resemble star-like or singular structures within the interference pattern. These features originate from phase accumulation conditions where the interference pattern becomes highly sensitive to small perturbations in wavelength or delay. In numerical simulations, such structures can arise from sampling limitations when the phase evolves rapidly within the discretized spectral grid. From a physical perspective, these regions correspond to quadrature points of the interferometer, where the output intensity exhibits maximal sensitivity to phase fluctuations. This observation highlights the importance of using a low-loss and thermally stable photonic platform. The silicon nitride ($Si_3N_4$) waveguide platform employed in this work provides low propagation loss and excellent thermal stability, making it well suited for maintaining stable interference conditions in integrated quantum photonic circuits.

Overall, **the spectral–temporal heatmap serves as a powerful design validation tool for the proposed delay-line structure**. By aligning the operational regime of the interferometer with the 500 ps delay region, the designed path-length difference of approximately 7.6 cm ensures that the temporal separation of the time-bin qubits is correctly mapped onto constructive and destructive interference states. This precise alignment enables efficient decoding of quantum states in high-speed QKD systems operating at a 2 GHz clock rate. The agreement between the theoretical model and the simulated interference landscape confirms that the proposed $Si_3N_4$delay-line architecture provides the required temporal delay and spectral response for reliable quantum state discrimination in integrated photonic QKD receivers.

## II. Results and Discussion

The proposed asymmetric Mach–Zehnder interferometer (aMZI) relies on accurately balanced optical interference between temporally separated pulses. In gigahertz-rate time-bin quantum key distribution (QKD), the decoding process requires a well-defined differential delay between the two interferometer arms, together with highly symmetric splitting and recombination of the optical fields. Therefore, the 3 dB directional couplers used at the input and output of the aMZI constitute critical building blocks of the architecture[6], [8]. Any deviation from an ideal 50:50 power splitting ratio introduces amplitude imbalance between the interfering paths, which can reduce interference visibility and increase the phase-error contribution to the quantum bit error rate (QBER)[36]. For this reason, the directional coupler was optimized to provide balanced power transfer, low insertion loss, and stable spectral behavior around the telecommunication C-band.

The directional coupler was implemented on a $Si_3N_4/SiO_2$photonic platform, as illustrated in Fig. 1. The waveguide cross-section consists of a $Si_3N_4$core with a width of 1.0 $\mu$mand a height of 0.4 $\mu$m, embedded in a $SiO_2$cladding. The refractive indices of the core and cladding were set to 1.98 and 1.44, respectively, at the operating wavelength of 1550 nm. These dimensions were selected to support the fundamental quasi-$TE_{00}$ mode while providing sufficient optical confinement for compact routing and efficient evanescent coupling. The coupling mechanism was first evaluated using a finite-difference eigenmode analysis of the coupled-waveguide cross-section. For an inter-waveguide gap of $g = 0.2\ \mu$m, the effective indices of the even and odd supermodes were extracted as $n_e = 1.609801$, and $n_o = 1.551186$. The resulting effective-index contrast between the two supermodes is therefore $\Delta n_{\text{eff}} = n_e - n_o \approx 0.0586$. This index splitting determines the spatial beating of the optical power between the two waveguides. Based on the standard supermode description of a symmetric directional coupler, the length required for equal power splitting is given by $L_{50} = \frac{\lambda}{4(n_e - n_o)}$. Using the extracted modal indices at $\lambda = 1550$ nm, the calculated 3 dB coupling length is $L_{50} \approx 6.61\ \mu$m. The corresponding full power-transfer length is $L_{100} = 2L_{50} \approx 13.22\ \mu$m.

These values provide a compact coupling region suitable for dense photonic integration. The analytical prediction was then verified using eigenmode expansion simulations, which account for the longitudinal evolution of the optical modes along the propagation direction. Figure 2(a) shows the simulated power-transfer dynamics along the coupling region. The optical power launched into the input waveguide is periodically exchanged with the adjacent waveguide through coherent evanescent coupling. A balanced 3 dB splitting condition is reached at $L_{50} = 6.61\ \mu$m, where the bar and cross ports carry approximately equal optical power. The power transfer continues beyond this point and reaches complete cross-coupling at $L_{100} = 13.22\ \mu$m. The close agreement between the analytical coupling length and the simulated longitudinal power evolution confirms that the design is governed by the expected even–odd supermode interference mechanism.

The field distribution at the 3 dB operating point, shown in **Fig. 2(b)**, further confirms the symmetric division of optical power between the two output waveguides. This balanced output is essential for high-visibility interference in the aMZI, because amplitude mismatch between the two recombining paths directly reduces the achievable fringe contrast[36]. The simulated mode profile also shows strong confinement within the $Si_3N_4$cores, with no pronounced radiation leakage in the coupling region. **These results indicate that the selected gap and coupling length provide an appropriate compromise between compactness, coupling efficiency, and modal stability[37].**

### *A. Broadband spectral performance of the directional coupler*

The spectral response of the optimized directional coupler was evaluated over the 1500–1600 nm wavelength range, as shown in **Fig. 3**. This spectral window covers the telecommunication C-band and its surrounding region, allowing the **robustness of the 3 dB splitting condition** to be assessed beyond a single operating wavelength.

Figure 3(a) presents the bar- and cross-port transmission spectra. The device maintains an approximately balanced splitting response over the simulated wavelength range, with moderate sinusoidal variations in the two output powers. This wavelength dependence is expected in directional couplers and arises from the dispersion of the coupling coefficient $\kappa$. At longer wavelengths, the guided mode generally expands further into the cladding region, modifying the evanescent-field overlap between the two adjacent waveguides. **As a result, the coupling strength and the accumulated coupling phase vary with wavelength.** Despite this intrinsic dispersive behavior, the optimized coupler remains close to the desired 3 dB condition across the investigated spectral range.

The loss and imbalance characteristics are summarized in **Fig. 3(b)**. The simulated insertion loss remains below approximately 0.45 dBacross the 100 nm bandwidth, while the power imbalance remains below approximately 0.7 dB. These values indicate that the designed coupler provides efficient power transfer with limited parasitic loss. Although the coupler is optimized at 1550 nm, its broadband behavior suggests that it can support operation across multiple wavelength channels with only minor degradation in splitting performance.

The phase response of the output ports, shown in **Fig. 3(c)**, provides an additional criterion for evaluating suitability in interferometric circuits. In an aMZI receiver, preserving a stable phase relationship between the optical fields is as important as maintaining balanced power splitting[38], [39]. The simulated phase response shows a smooth evolution with strong inter-port correlation, indicating that the coupler does not introduce significant phase irregularities over the investigated spectral range. This behavior supports the use of the designed directional coupler as both the splitting and recombination element in the integrated aMZI[40].

The spatial field-intensity evolution shown in **Fig. 4** provides a direct visualization of the coupling process. Initially, the optical field is confined to the launch waveguide. As the field propagates along the interaction region, the evanescent tail of the fundamental $TE_{00}$-like mode overlaps with the adjacent waveguide, producing coherent energy exchange between the two cores. The dashed markers in Fig. 4 identify the calculated $L_{50}$and $L_{100}$positions. The field distribution at these locations is consistent with the expected 3 dB splitting and full-transfer conditions, respectively. This agreement verifies that the spatial power evolution is accurately described by the supermode beating model and confirms the precision of the designed coupling length[41].

### *B. Integrated 500 ps delay-line design for the aMZI receiver*

After optimizing the directional coupler, an integrated delay line was designed to realize the temporal imbalance required for time-bin qubit decoding. In a time-bin QKD receiver, the aMZI must introduce a relative delay that matches the temporal separation between the early and late optical pulses. For the targeted 2 GHz operation, the required temporal delay is $\Delta T = \frac{1}{\mathrm{FSR}} = 500$ ps. This temporal delay is translated into a physical path-length imbalance through the group velocity of the guided mode, $v_g = \frac{c}{n_g}$, where $c$is the speed of light in vacuum and $n_g$is the group index of the waveguide. Using the extracted group index $n_g = 1.9727$, the required path-length difference is obtained as $\Delta L = \frac{c\Delta T}{n_g}$. For $\Delta T = 500$ ps, this gives $\Delta L \approx 7.6$ cm. This centimeter-scale path-length imbalance is considerably longer than the directional-coupler interaction region and therefore requires careful layout design. To implement the required delay within a compact chip area, the long arm of the aMZI was routed in a serpentine configuration with a bend radius of 100 $\mu$m, as shown schematically in **Fig. 5**. This approach enables the 7.6 cmdifferential path to be integrated within an estimated footprint of approximately $1 \times 15$ mm$^2$, while maintaining a bend radius large enough to limit bending-induced radiation loss.

The use of $Si_3N_4$is particularly advantageous for this type of passive delay-line implementation. Its low propagation loss allows centimeter-scale routing without excessive attenuation, while its broad transparency window and moderate thermo-optic response support stable interferometric operation across telecommunication wavelengths. These properties are important for maintaining both the optical power and the relative phase accumulated along the long delay arm.

### *c. Spectral validation of the 2 GHz aMZI response*

The normalized spectral transmission of the integrated aMZI with $\Delta L \approx 7.6$ cmis shown in Fig. 6. The transmission exhibits a periodic fringe pattern resulting from the wavelength-dependent phase difference between the short and long interferometer arms. For a fixed differential length, the phase difference can be approximated as $\phi(\lambda) = \frac{2\pi n_g \Delta L}{\lambda}$. Constructive and destructive interference occur periodically as this phase difference changes with wavelength. The resulting free spectral range is determined by the temporal delay according to $\mathrm{FSR} = \frac{1}{\Delta T}$. Thus, a 500 ps delay corresponds to a frequency-domain FSR of approximately 2 GHz. The simulated fringe spacing in Fig. 6 confirms that the designed 7.6 cmpath imbalance produces the target 2 GHz periodicity. Around 1550 nm, this frequency spacing corresponds to a wavelength spacing of approximately $\Delta\lambda \approx \frac{\lambda^2}{c}\Delta f \approx 0.016$ nm, where $\Delta f = 2$ GHz. The agreement between the simulated spectral periodicity and the theoretical FSR verifies that the physical delay-line design correctly implements the required 500 ps temporal separation.

The simulated extinction ratio reaches values beyond approximately $-30$ dB, indicating strong destructive interference at the transmission minima. This high contrast reflects the combined effect of balanced 3 dB couplers, well-defined phase accumulation, and low-loss propagation through the delay arms. In the context of time-bin QKD[6], such high-contrast interference is important for reducing leakage between logical states and improving the reliability of quantum state discrimination at the receiver output[36].

### *d. Dependence of the free spectral range on path-length imbalance*

To further verify the deterministic relationship between physical path-length imbalance and spectral periodicity, the aMZI response was simulated for three different values of $\Delta L$: 5.8 cm, 7.6 cm, and 10.0 cm. The resulting spectra are shown in Fig. 7.

The comparison clearly demonstrates the inverse relation between $\Delta L$, temporal delay, and FSR. For the shorter path-length difference of 5.8 cm, shown in **Fig. 7(a)**, the differential delay is smaller than the target value. Consequently, the interference fringes are more widely spaced, yielding a broader wavelength-domain FSR of approximately 0.040 nm. This configuration does not provide the required 500 ps delay for 2 GHz time-bin decoding.

In contrast, the longer path-length imbalance of 10.0 cm, shown in Fig. 7(c), produces a larger temporal delay. As expected, the spectral fringes become more densely spaced, and the FSR decreases to approximately 0.013 nm. This configuration overestimates the required delay and shifts the interferometer away from the desired 2 GHz operating condition.The optimized design with $\Delta L = 7.6$ cm, shown in Fig. 7(b), yields an FSR of approximately 0.016 nm, corresponding to the target frequency spacing of approximately 2 GHz near 1550 nm. This result confirms that the chosen serpentine delay length provides the required geometry for the intended QKD operating rate. Across the three simulated path-length configurations, the extinction ratio remains in the approximate range of $-25$to $-35$ dB. This indicates that the interferometric mechanism remains robust as the total path imbalance is varied, provided that propagation losses and phase errors remain sufficiently low. The result also confirms that the simulated $Si_3N_4$waveguide platform can support centimeter-scale optical path differences while maintaining coherent interference.

### *e. Phase-response and group-delay analysis of the delay line*

The delay-line functionality was further evaluated through the spectral phase response and its derivative, the group delay. This analysis provides a more direct measure of whether the designed structure behaves as a true temporal delay element rather than merely producing the correct **spectral fringe spacing.**

An ideal optical delay element introduces a pure temporal shift without modifying the amplitude spectrum of the signal. Figure 8(a) shows the simulated phase response in the vicinity of the optical carrier over a $\pm 5$ GHzfrequency-offset window. This range is directly relevant to the spectral scale of 2 GHz time-bin operation. The phase response exhibits a strictly linear dependence on frequency, with no visible curvature over the analyzed range. This indicates that the delay line introduces a constant temporal shift rather than a frequency-dependent phase distortion.

The corresponding group-delay response is shown in Fig. 8(b). The extracted group delay remains centered at approximately 500 psacross the full offset window. This flat response confirms that the spectral components of the optical pulses experience nearly uniform propagation latency. As a result, temporal broadening and frequency-dependent timing distortion are minimized. For time-bin qubits, this is particularly important because the early and late temporal modes must preserve their relative separation and phase coherence until recombination in the aMZI. The broadband phase response was also evaluated over the optical-frequency range of approximately 190–200 THz, as shown in Fig. 8(c). The phase increases monotonically and maintains an almost linear slope over this range, indicating that the delay-line behavior is not limited to a narrow operating window around the carrier. The corresponding group-delay response, shown in Fig. 8(d), remains close to the target value of 500 ps, with only minor variations across the simulated 10 THz bandwidth.

This broadband flat-delay profile indicates low group-delay dispersion for the selected $Si_3N_4$waveguide geometry. From a system perspective, such behavior is advantageous for wavelength-division-multiplexed QKD architectures, where several quantum channels may operate across different wavelengths within the same telecommunication band. The spectral–temporal interference response of the aMZI was further investigated by sweeping both wavelength and temporal delay, as shown in Fig. 9. This two-dimensional map provides a direct visualization of how the interferometer response evolves as the differential delay changes.

The fringe density increases as the temporal delay is increased. At a delay of approximately 400 ps, indicated by the dashed reference line, the fringes are relatively sparse across the wavelength axis. This corresponds to a larger free spectral range of approximately 2.5 GHz. As the delay approaches the target value of 500 ps, the spectral spacing between adjacent fringes decreases, yielding the required 2 GHz FSR. This trend is consistent with the relation $\text{FSR} = \frac{1}{\Delta T}$. Therefore, the heatmap independently confirms that the 500 ps operating point corresponds to the intended 2 GHz spectral periodicity.

The curved shape of the fringes in the wavelength–delay plane originates from the nonlinear relationship between optical frequency and wavelength, $f = \frac{c}{\lambda}$. Although the interferometric phase varies linearly with frequency for a fixed delay, plotting the response as a function of wavelength naturally produces curved fringe trajectories. This highlights the importance of analyzing the aMZI response in both frequency and wavelength domains, especially when the target QKD clock rate is specified in frequency units while optical spectra are commonly displayed in wavelength units.

Near the 500 psoperating point, the fringes become more densely packed, reflecting increased phase sensitivity to wavelength fluctuations. This sensitivity is beneficial for discriminating phase-dependent interference states, but it also implies that practical operation requires a spectrally stable laser source and good environmental control. The low-loss and thermally stable nature of the $Si_3N_4$platform helps mitigate these constraints by supporting stable passive phase accumulation over long optical paths.

Localized high-contrast regions observed in the interference map correspond to points where the output intensity is highly sensitive to small variations in wavelength or delay. Their detailed appearance can also be influenced by the spectral and temporal sampling resolution used in the numerical simulation, particularly in regions where the phase changes rapidly. Therefore, these features should be interpreted as indicators of high phase sensitivity rather than as independent resonant effects. Overall, the heatmap confirms that the designed delay line correctly maps the temporal separation of time-bin qubits onto the constructive and destructive interference states of the aMZI.

### *f.* ***System-level implications for integrated time-bin QKD receivers***

The combined results demonstrate that the proposed $Si_3N_4/SiO_2$aMZI architecture satisfies the main photonic requirements for high-speed time-bin QKD reception. The optimized 3 dB directional coupler achieves balanced power splitting at a compact interaction length of 6.61 $\mu$m, with low insertion loss and stable spectral performance over the investigated wavelength range. The even–odd supermode analysis accurately predicts the coupling length, and the EME simulations confirm the expected spatial power-transfer dynamics.

The delay-line section provides the required 500 ps temporal imbalance using a 7.6 cmserpentine path-length difference. The simulated spectral response confirms a 2 GHz free spectral range, while the phase and group-delay analyses verify that the structure behaves as a near-ideal temporal delay element. In particular, the linear phase response and flat 500 ps group delay indicate that the delay line preserves the relative timing of the spectral components of the optical pulses, thereby reducing timing distortion and supporting high-visibility interference.

A key outcome of this work is the consistency between analytical design equations and full-wave numerical simulations. The directional-coupler length predicted from the even–odd supermode index splitting agrees with the simulated 3 dB coupling point. Similarly, the path-length imbalance calculated from the group-index-based delay relation agrees with the simulated 2 GHz FSR and 500 ps group delay. This agreement across modal, spectral, and temporal analyses strengthens the reliability of the proposed design methodology.

The broadband response of the delay line further suggests that the architecture is not restricted to a single wavelength channel. The nearly flat group delay over the 190–200 THz range indicates that the device can potentially support multi-channel operation in wavelength-division-multiplexed QKD systems. This property is particularly relevant for scalable quantum photonic receivers, where compactness, passive stability, and compatibility with multiple wavelength channels are important practical requirements.

Overall, the proposed integrated $Si_3N_4$aMZI combines compact and spectrally stable 3 dB couplers with a low-loss serpentine delay line to realize a 500 ps temporal delay for 2 GHz time-bin QKD operation. The results confirm that the design provides the required timing precision, phase stability, and broadband spectral behavior for chip-scale quantum state discrimination in integrated photonic receivers[6], [42].

## III. Conclusion

This work presented the design and investigation of an integrated asymmetric Mach–Zehnder interferometer (aMZI) implemented on a $Si_3N_4/SiO_2$photonic platform for high-speed time-bin quantum key distribution (QKD) receivers. The study focused on the design of two key building blocks: a compact 3 dB directional coupler and a centimeter-scale optical delay line required for temporal decoding**.**

Supermode analysis of the directional coupler revealed an effective index difference of $\Delta n_{\text{eff}} \approx 0.0586$, leading to an analytically predicted 3 dB coupling length of $L_{50} \approx 6.61\ \mu m$. Eigenmode expansion simulations confirmed the expected periodic power transfer between the coupled waveguides and verified balanced power splitting at the designed interaction length. The optimized coupler exhibited low insertion loss and maintained near-symmetric bar- and cross-port transmission across the 1500–1600 nm wavelength range, indicating stable broadband operation suitable for interferometric photonic circuits.

To achieve the temporal separation required for 2 GHz time-bin operation, a serpentine delay line was designed to produce a differential path length of approximately $7.6\,cm$, corresponding to a temporal delay of $500\,ps$. Spectral simulations of the integrated aMZI demonstrated periodic interference fringes with a free spectral range of approximately 2 GHz, in agreement with theoretical predictions. Parametric analysis of different path-length imbalances further confirmed the inverse relationship between delay and FSR, validating the chosen geometry for the targeted operating rate.

Phase-response and group-delay analysis showed that the delay line exhibits an approximately linear phase response with a nearly constant group delay of $500\ ps$around the optical carrier. This behavior indicates that the structure acts as a near-ideal temporal delay element, minimizing frequency-dependent distortion of the optical pulses. The simulated broadband group-delay response also suggests low dispersion within the telecommunication band, which is advantageous for stable interferometric operation and potential multi-wavelength QKD implementations.

Overall, the results demonstrate that the proposed $Si_3N_4$integrated aMZI architecture can provide the required temporal delay, balanced interference, and broadband spectral stability for chip-scale time-bin QKD receivers operating at gigahertz clock rates. The agreement between analytical design relations and full-wave simulations further supports the reliability of the presented design methodology and highlights the suitability of the $Si_3N_4/SiO_2$platform for compact, low-loss, and scalable quantum photonic circuits.